\documentclass[sigconf,nonacm]{acmart}
\usepackage{multirow}
\usepackage{pdfpages}

\AtBeginDocument{%
  \providecommand\BibTeX{{%
    \normalfont B\kern-0.5em{\scshape i\kern-0.25em b}\kern-0.8em\TeX}}}

\begin{document}

\title[A Study on Hyperparameters for a Self-Adaptive HAR System]{A Study on Hyperparameters Configurations for an Efficient Human Activity Recognition System}
\subtitle{Extended Version}
\titlenote{This paper, followed by Appendix~\ref{ext} with a technical report with extensive content and results, is an extended version of the paper:
Paulo J.S. Ferreira, João Mendes-Moreira, and João M.P. Cardoso. 2023. A
Study on Hyperparameters Configurations for an Efficient Human Activity
Recognition System. In 8th International Workshop on Sensor-Based Activity Recognition and Artificial Intelligence (iWOAR 2023), Sept. 21–22, 2023, Lübeck, Germany. ACM, New York, NY, USA. \url{https://doi.org/10.1145/3615834.3615851}}

\author{Paulo J.S. Ferreira}
\authornote{All authors contributed equally to this research.}
\email{up201305617@fe.up.pt}
\orcid{0000-0001-6553-4220}
\author{João Mendes-Moreira}
\orcid{0000-0002-2471-2833}
\email{jmoreira@fe.up.pt}
\authornotemark[1]
\author{João M.P. Cardoso}
\orcid{0000-0002-7353-1799}
\email{jmpc@fe.up.pt}
\authornotemark[1]
\affiliation{%
  \institution{INESC TEC, Faculty of Engineering, University of Porto}
  \streetaddress{R. Dr. Roberto Frias s/n }
  \city{Porto}
  \country{Portugal}
  \postcode{4200-465}
}

\renewcommand{\shortauthors}{Ferreira, et al.}

\begin{abstract}
Human Activity Recognition (HAR) has been a popular research field due to the widespread of devices with sensors and computational power (e.g., smartphones and smartwatches). Applications for HAR systems have been extensively researched in recent literature, mainly due to the benefits of improving quality of life in areas like health and fitness monitoring. However, since persons have different motion patterns when performing physical activities, a HAR system must adapt to user characteristics to maintain or improve accuracy. Mobile devices, such as smartphones, used to implement HAR systems, have limited resources (e.g., battery life). They also have difficulty adapting to the device’s constraints to work efficiently for long periods. In this work, we present a \textit{k}NN-based HAR system and an extensive study of the influence of hyperparameters (window size, overlap, distance function, and the value of k) and parameters (sampling frequency) on the system accuracy, energy consumption, and inference time. We also study how hyperparameter configurations affect the model's user and activity performance. Experimental results show that adapting the hyperparameters makes it possible to adjust the system's behavior to the user, the device, and the target service. These results motivate the development of a HAR system capable of automatically adapting the hyperparameters for the user, the device, and the service.
\end{abstract}

\ccsdesc[500]{Human-centered computing~Mobile devices}

\keywords{human activity recognition, HAR, \textit{k}NN, energy consumption, adaptability, smartphones, hyperparameters, PAMAP2, Pareto-Front, sampling frequency}



\maketitle

\section{Introduction}
Recently, Human Activity Recognition (HAR)~\cite{computers9040096,sports,mot_inc_1,sign,Gupta2021} has been a popular research field due to the wide spread of devices with sensors (e.g., smartphones and smartwatches). HAR aims at recognizing the activities performed by humans at runtime through the analyses of sensing data and other observations acquired by in situ devices ~\cite{har_def}. HAR has enabled novel applications in different areas, such as health and fitness monitoring, security, smart cities and entertainment ~\cite{shoaib,xing,overview}. 

\begin{figure}[ht]
  \centering
  \includegraphics[width=\linewidth]{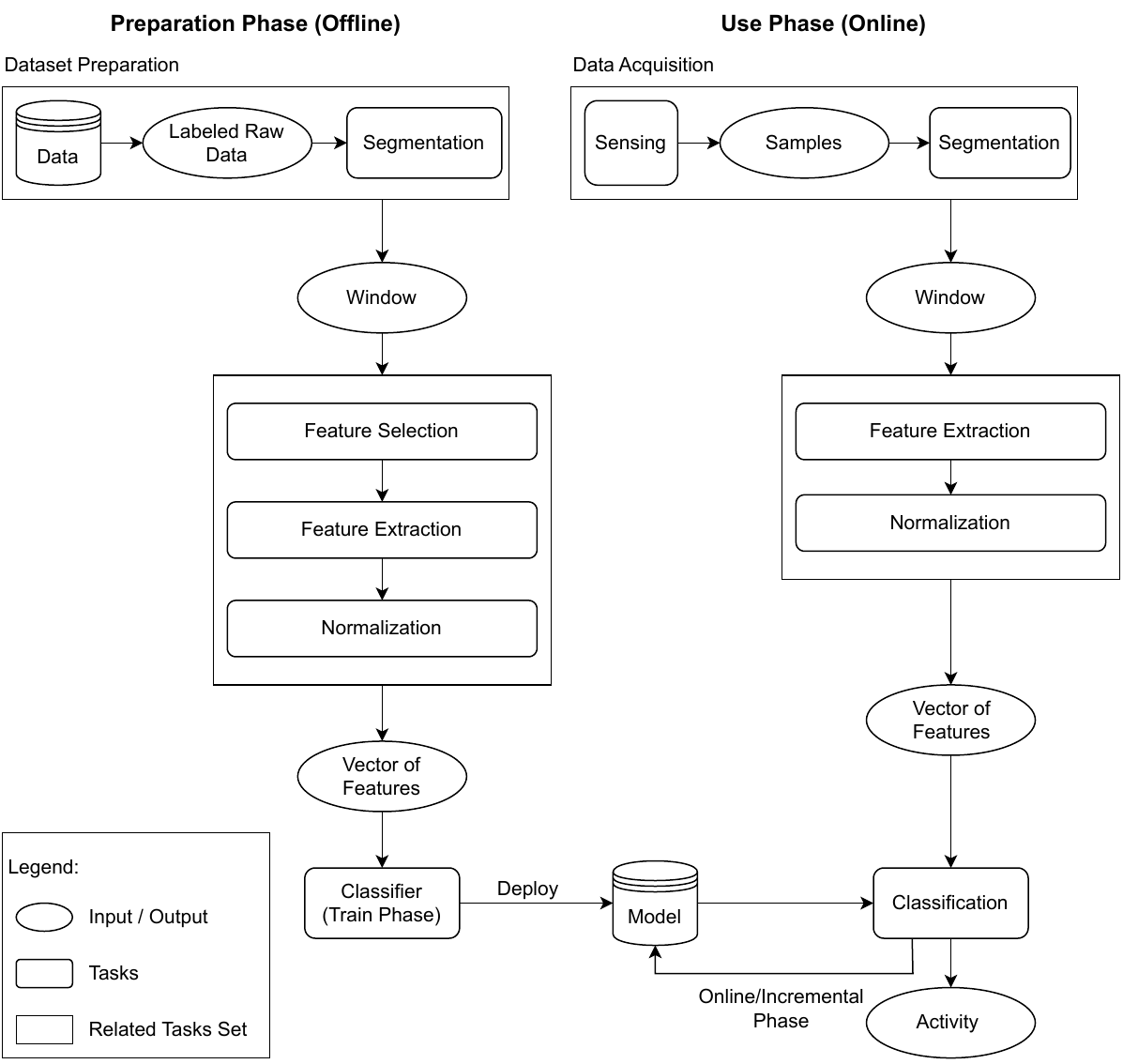}
  \caption{Block diagram of the HAR system considered.}
  \Description{A block diagram of a typical HAR system implemented on a mobile device.}
  \label{fig:harsystem}
\end{figure}

Figure~\ref{fig:harsystem} shows the main stages of the two views of a ML-based HAR system: the offline training view (Preparation Phase), and the online classification view (with the possibility of incremental/online learning) (Use Phase).

The Preparation Phase is used to train the model to be later used in the classification process. This Phase includes the following steps: data acquisition, pre-processing, and training~\cite{shoaib}. At this Phase, after data acquisition, data is labeled and stored in files to build the dataset used in the system. In the Use Phase, the trained classifiers infer activities. This step can be done online on a mobile device. This Phase includes the following stages: data acquisition, pre-processing, and classification~\cite{shoaib}. This Phase can also contain the Incremental Learning step. In this case, data are not stored in files during data acquisition online. Instead, it goes directly to the segmentation. The feature selection is only carried out in the pre-processing step of the Preparation Phase.

Smartphones have been increasingly used to develop HAR systems, due to the capability and diversity of the sensors embedded in these devices~\cite{overview}. The advantage of smartphones over other wearable devices is associated with (a) their widespread use, (b) their ability to collect and process sensor data~\cite{overview}, (c) the increase in their processing power, (d) the increase in their battery capacity~\cite{shoaib}, and (e) their use by most people. Although smartphones have seen rapid development in recent years, these devices continue to have limitations concerning battery life and available memory. 

In order to be able to recognize different activities accurately, a HAR system uses Machine Learning (ML) algorithms capable of inferring activities from the sensor data. One of the most used ML algorithms in HAR is \textit{k}-Nearest Neighbours (\textit{k}NN)~\cite{kNN}. \textit{k}NN is an instance-based lazy learning algorithm. This fact facilitates the implementation of incremental/online learning HAR systems in mobile devices since \textit{k}NN does not require a computationally expensive training stage. Besides, \textit{k}NN has also been proved to obtain very high accuracy in the HAR field (see, e.g., ~\cite{acc_4,kose2012online,knn_1,knn_2,knn_3,Celik2022,Mubibya2022,Xia2022,computers9040096}). Using \textit{k}NN, the implementation of online/incremental learning approaches only needs to add, remove or update the instances in the \textit{k}NN memory. Updating the \textit{k}NN is fast and requires low energy because there is no need to retrain the algorithm, which is essential in devices with energy limitations~\cite{computers9040096}.  

Typically, an ML algorithm has two types of parameters: model parameters and hyperparameters~\cite{luo2016review}. Model Parameters are used inside the model and are estimated or learned from the data as a part of the learning process. On the other hand, hyperparameters are external parameters that are not part of the model and thus can not be predicted from the data set but can be configured by subject matter experts or by trial and error until an acceptable accuracy is achieved. Hyperparameters are used to control the learning process and can have wildly varying effects on the resulting model and on its performance~\cite{hyper}. For the same training dataset, with different hyperparameters, an ML algorithm might learn models that have significantly different performances on the testing dataset~\cite{hyper_2}. 


This paper studies the influence of hyperparameters on a HAR system's optimal performance, specifically analyzing window size, overlap, \textit{k}value, distance function, and sampling frequency in a \textit{k}NN-based approach. Accuracy, response time, and energy consumption are evaluated as performance metrics. While most studies focus on hyperparameters' impact on accuracy, limited research explores their influence on critical measures such as response time and energy consumption. In real-world applications, these factors significantly impact the practicality and usability of HAR systems. Prompt response time is crucial for time-sensitive applications that require real-time activity recognition, while energy consumption is vital for mobile and wearable devices with limited battery life. 

Moreover, HAR systems need to be adaptable to changing conditions, such as different users or environments. To achieve adaptability, it is essential to study the hyperparameter's impact on a HAR system's performance under different scenarios. For instance, a HAR system optimized for one user may not perform well for another user due to differences in their physical characteristics or activity patterns. Similarly, a HAR system designed for indoor environments may not perform well in outdoor environments due to changes in lighting, temperature, and other factors.

Therefore, in this study, we address the impact of hyperparameters on the performance of a HAR system in terms of accuracy, response time, and energy consumption. Specifically, we explore the effect of different hyperparameter configurations on the performance of a HAR system using the PAMAP2~\cite{pamap,pamap2} dataset. PAMAP2 is one of the most widely used wearable sensor-based datasets in HAR research and also has been widely used to test and implement new approaches for HAR systems~\cite{computers9040096,Yuxun2023,Mubibya2022,Mohammed2023,Trinh2022,Huang2022,Cheng2022,Yang2022}. This dataset has data from different sensors, body placements, and users. 
The results of this study provide valuable insights into the optimal hyperparameter configurations for HAR systems, enabling the design of more efficient, effective, and adaptable systems for real-world applications. We present in~\cite{anon} an extended version of this study.

Specifically, in this study, we answer the following Research Questions:

\begin{itemize}
    \item RQ 1 - How do hyperparameter configurations affect the accuracy, response time, and energy consumption of a HAR system globally and across different users and contexts?
    \item RQ 2 - What are the overall, and specific activities, the impact of window size, overlap, distance function, and \textit{k}for a \textit{k}NN-based HAR system regarding its accuracy, response time, and energy consumption? 
    \item RQ 3 - How do the sampling frequencies influence the accuracy of a HAR system for different users and activities? 
    \item RQ 4 - How does reducing the sampling frequency change the trade-off between accuracy and energy consumption for different hyperparameter configurations of a HAR system for a specific user?
    \item RQ 5 - How does the mismatch between the sampling frequencies of the train and test sets affect the accuracy of a HAR system?
\end{itemize}

The remainder of this paper is organized as follows. Section~\ref{related} describes the related work regarding the study of hyperparameters. Section~\ref{setup} describes the experimental setup and methodology used to conduct this study. Section~\ref{results} presents and discusses the experimental results, including a comparison with the Related Work. Finally, in Section ~\ref{conclusion}, we present the main conclusions for this study and the future work planned.

\section{Related Work} \label{related}


HAR is a research field focused on automatically identifying and classifying human activities from sensor data. ML algorithms are commonly used in HAR systems for activity classification. Hyperparameters are essential for building accurate HAR models. They can influence the model's performance significantly, so choosing them carefully is important. Hyperparameters affect the algorithms' performance in terms of prediction and computation, so many researchers are investigating their impact on HAR systems. The most common hyperparameters in traditional ML algorithms are window size and overlap. These hyperparameters are used to segment the data for both Preparation and User Phases (see Figure 1). Here are some examples of HAR studies that examined some hyperparameters.

Banos et al.~\cite{Banos2014}, uses the REALDISP~\cite{realdisp} dataset to evaluate the impact of the window size, using C4.5, \textit{k}NN, Naïve Bayes, and Nearest Centroid Classifier as inference algorithms. The dataset has data for 33 different activities and was collected at 50 Hz. To measure the performance, the authors use F1-Score. They use windows sizes between 0.25 s and 7 s. Overall, they show that the windows with 2 s achieve the best results. They also concluded that larger window sizes not always translated into better performances for the models.  


Garcia et al. study the impact of window size and overlapping. They used the PAMAP2 HAR dataset and built an ensemble classifier with \textit{k}NN, VFDT, and Naive Bayes algorithms. The study explored variations in window size (from 100 to 1000 samples with increments of 100) and overlap (from 0.0 to 0.9 with increments of 0.1). Evaluation metrics included accuracy, energy consumption, and execution time. They use the ODROID-XU+E6 board to measure energy consumption. Smaller window sizes showed lower accuracy, while larger sizes improved accuracy. The overlap factor had fluctuations in accuracy, with optimal results between 0.1 and 0.5. Smaller windows consumed less energy and had shorter execution times due to reduced feature calculation effort. Increasing the overlap factor raised energy consumption due to more calculations and classifications. Higher accuracies were associated with increased energy consumption.


Wang et al.~\cite{wang_2018} conducted a study on the impact of window size in a HAR system. They used a private dataset containing activities such as still, walking, up/down stairs, up/down an elevator, up/down escalator, typing, swinging, phoning, and trouser pocket. The data was collected at a sampling frequency of 50 Hz. The authors varied the window size between 0.5 s and 7 s. They evaluated the performance of different algorithms including SVM, \textit{k}NN, Decision Tree, Naïve Bayes, and Adaboost, using the F1-Score metric. The results showed that increasing the window size significantly improved the classification performance for all algorithms. However, larger window sizes also resulted in increased recognition latency. Based on their findings, the authors concluded that a window size between 2.5-3.5 s provided the best tradeoff between performance and latency.

Although the window size and overlap are among the most studied hyperparameters, they control data segmentation before being used to train or use the traditional ML algorithms. They are not specific to any ML algorithm. However, traditional ML algorithms have specific hyperparameters (e.g., \textit{k}NN has the number of \textit{k}, distance function, and the maximum number of instances). These hyperparameters are defined in the Preparation Phase, more precisely in the training stage of the ML algorithm (see Figure~\ref{fig:harsystem}). Some authors are also focusing on studying the algorithm's specific hyperparameters.

Mohsen et al.~\cite{Mohsen2022} study the impact of the value of \textit{k}on model accuracy, varying \textit{k}from 1 to 20 and using the UCI-HAR dataset. The results showed that accuracy improved as \textit{k}increased, with the best results achieved when \textit{k}was between 8 and 20. However, no significant improvement in accuracy was observed for \textit{k}values greater than 8. Liu et al.~\cite{Liu2021} also assessed the effect of different values of \textit{k}on HAR datasets, specifically on HAPT and Smartphone Datasets for Human Activity Recognition in Ambient Assisted Living Data (Smartphone). They examined \textit{k}values of 3 and 9 in terms of accuracy. The findings demonstrated that increasing the value of \textit{k}enhanced model accuracy. However, for \textit{k}values greater than 6, accuracy started to decrease. The optimal value of \textit{k}for both datasets was determined to be 6, as it yielded the highest accuracy.

Sampling frequency, although not considered a hyperparameter, plays a crucial role in the performance of a Human Activity Recognition (HAR) system. It directly affects inference time, energy consumption, and the performance of machine learning algorithms. Higher sampling frequencies, meaning more samples per second, generally increase energy consumption. Several studies have investigated the impact of sampling frequency on HAR systems.

Santoyo-Ramón et al.~\cite{Ramon2022} study the influence of the sampling frequency in 15 HAR datasets. The datasets have sampling frequencies between 18 Hz and 238 Hz. The metrics used are sensitivity, specificity, and the geometric mean as single descriptors of the global system performance. They use a convolutional neural network (CNN) as an inference algorithm. The results show that the system performance only degrades when the sampling rate of the dataset is between 10 and 15 Hz. The inference improves when the sampling frequency of the dataset is between 20 and 40 Hz. 


Niazi et al.~\cite{Niazi2017} conducted a study on the effects of sampling frequency and window duration in a HAR system using a private dataset collected at 100 Hz. Window sizes of 1, 2, 3, 5, and 10s, along with sampling frequencies of 5, 10, 20, 25, 50, and 100 Hz, were considered. Random Forest was used as an inference algorithm. Statistical analysis with weighted least squares and two-way factorial ANOVA was performed to calculate the expected average value (EV) for each combination of window duration and sampling frequency. The optimal combination was found to be 10s/50Hz, especially for young, physically active individuals. Higher sampling rates and window sizes showed higher significance in the EVs, but reducing the sampling rate to 20Hz did not significantly affect accuracy.

Zheng et al.~\cite{Zheng2017} evaluate the impact of the sampling frequency on accuracy and energy consumption with SVM as an inference algorithm. They use a dataset at 1 HZ, 5Hz, 10 Hz, and 50 Hz sampling frequencies. The data consist of the following activities: Sitting, Standing, Walking, Running, Upstairs, and Downstairs. The results show that the accuracy has only improved slightly with the sampling rate increase from 1 Hz to 50 Hz. In terms of energy consumption, there is an increase with increased sampling frequency, which is more significant when the sampling rate changes from 10 Hz to 50 Hz. 

In these studies, the impact of the hyperparameters is more focused on the performance of the resulting model instead of their influence on the computational cost of the system. However, some exceptions exist, like Garcia et al.~\cite{kemilly}, who, in addition to accuracy, also evaluate the energy consumption for different window sizes and overlap values. Unlike the studies presented in this section, our study considers 5 different hyperparameters of different stages of a HAR system (window size and overlap for data segmentation, sampling frequency for data acquisition, and distance function and the \textit{k}value for the training of the algorithm) and evaluate their impact on the performance of the model (accuracy and F1-Score) and the computational cost of the system (inference time and energy consumption). 

\section{Experimental Setup and Methodology} \label{setup}
This section presents the experimental setup and experiments carried out in this study. It includes the dataset and algorithm used in the experiments, the evaluation metrics, the evaluation procedure, the processing of the raw data (feature extraction and normalization), the running platforms and the experiments carried out in our study, and some procedures we used to prepare the data for the experiments. 

\textbf{Dataset:}
The \emph{PAMAP2}~\cite{pamap,pamap2} HAR dataset contains $1,926,896$ samples of raw sensor data from $9$ different users and $12$ different activities. Data were collected from $3$ Inertial Measurement Units (IMU) positioned in different body areas (wrist, chest, and ankle), at a sampling frequency of $100$ Hz, and a Heart-Rate Monitor at a sampling frequency of $9$ Hz. Each IMU has $4$ embedded sensors: a 3-axis accelerometer, a 3-axis gyroscope, a 3-axis magnetometer, and a thermometer. The activities are organized as basic activities (walking, running, nordic walking, and cycling); posture activities (lying, sitting, and standing); everyday activities (ascending and descending stairs); household (ironing and vacuum cleaning), and fitness activities (rope jumping). We decided to use only the following sensors: accelerometer, gyroscope, and magnetometer.

\textbf{Evaluation:}
To measure the performance of the models obtained using the explored hyperparameters, we use the following ML metrics: accuracy and F1-Score. We also measure the impact of different configurations on the system regarding response time (equals inference time, in our experiments) and energy consumption. Both response times and energy consumption refer to the time and energy needed for each inference. The values for time and energy include the following phases: reading data from the files, feature extraction, and inference. We use the fANOVA (Functional ANOVA (Analysis of Variance))~\footnote{https://www.automl.org/ixautoml/fanova/}~\cite{fanova} algorithm to assess the importance of the selected hyperparameters.
The experiments in this paper follow the leave-one-subject-out (LOSO) approach, where one user is selected for testing the model while the remaining users are used for training. This approach enables us to evaluate the impact of the same set of hyperparameters on different models and highlights the significance of a user-adaptive model. 





\textbf{Feature Extraction:}
 Using fixed-size sliding windows, we extracted $10$ features for each 3D sensor: x-axis Mean, y-axis Mean, z-axis Mean, Mean of the sum of the x, y, and z axes, x-axis Standard Deviation, y-axis Standard Deviation, z-axis Standard Deviation, x and y axes Correlation, x, and z axes Correlation, and y and z axes Correlation. The features are extracted from 9 sensors: 3 sensors (accelerometer, gyroscope, and magnetometer) for each body placement (wrist, chest, and ankle). This results in vectors of $90$ features that are the input to the ML algorithm. The features are normalized using the Min-Max technique.


\textbf{Inference ML Algorithm:}
\textit{k}-Nearest Neighbour (\textit{k}NN)~\cite{kNN} is an instance-based classifier based on the majority voting of its \textit{k} neighbours~\cite{xing} for classifying an instance. The value of \textit{k}defines how many nearest neighbor instances contribute to the classification of each instance. \textit{k}NN does not use any model to fit and it is only based on memory. \textit{k}NN is a lazy learning algorithm because it does not have a learning phase; instead, it "memorizes" the training dataset. The more the number of training instances stored in the \textit{k}NN's memory, the greater the number of distances to be calculated and, consequently, the longer the execution time of the inference phase. Usually, Euclidean distance is used as the distance metric ~\cite{knn_1}. The selection of \textit{k} is an important consideration as it affects the classification performance. 
In this study, the \textit{k}NN implementation is based on the MOA Java library. The implementation offers three hyperparameters that can be optimized: \textit{k} value, distance function, and the maximum number of instances that \textit{k}NN can store.

\textbf{Running Platforms:}
We ran the experiments on two platforms: a desktop computer and a board with ARM CPUs. The desktop computer was used to run the exploration experiments. The application was written in Java code and executed with version 1.8 of the Java Runtime Environment (JRE).
The experiments related to the HAR system run on an embedded Odroid computing board. They were used to measure energy consumption and response time. It represents a possible mobile/embedded device. We use an ODROID-XU+E6 \footnote{https://www.hardkernel.com/} equipped with an Exynos5 Octa CPU (used in several Samsung smartphones) with four big cores (ARM Cortex-A15 up to 1.6 GHz) and four small cores (ARM Cortex-A7 up to 1.2 GHz). ODROID-XU+E6 has four current/voltage sensors to measure the power consumption of the Big A15 cores, Little A7 cores, GPU and DRAM individually. In our experiments, we only consider the energy consumption for the A15 cores, A7 cores, and DRAM, as we are not using the GPU.


\textbf{Configurations Search:} \label{config_exp}
The goal of the first experiment was to explore the configuration space. We use the multi-objective search in Optuna~\cite{optuna_2019} to select the configurations that optimize two different objectives: maximize the accuracy and 
minimize response time. We select the hyperparameters window size, overlap, distance function, and the value of k. Table~\ref{tab:selec_hyper_1} shows the hyperparameters and their search space. To search for the best configurations, we conduct 1000 (out of 5400) trials in Optuna. Each trial represents a different configuration for the hyperparameters. However, due to the characteristics of the sampler used by Optuna (NSGAIISampler) to select the values, it was only possible to obtain 702 different configurations, i.e., 13\% of the total number of possible configurations. We split the dataset into training and validation sets to conduct this experiment.

\begin{table}[ht]
\centering
\caption{Selected Hyperparameters for the 1st exploration.}
\label{tab:selec_hyper_1}
\begin{tabular}{lll}
\toprule
\textbf{Hyperparameters}    & Search Space   & \# Values \\ \midrule
k          	                & [1; 10]  & 10\\ 
distance function           & \begin{tabular}[c]{@{}l@{}}Euclidean, Manhattan, \\ Chebyshev\end{tabular} & 3\\ 
window size	                & [50; 900] with steps of 50 & 18\\ 
overlap                     & [0.0; 0.9] with steps of 0.1 & 10\\ \midrule
\# Configurations           & & 5400 \\ \bottomrule
\end{tabular}
\end{table}


\textbf{Reducing the Number of Training Windows:}
One important aspect to consider is that the window size and overlap hyperparameters can affect the number of instances obtained from a dataset. To ensure a fair comparison, it matters to establish a consistent number of instances across all possible combinations of window size and overlap. To achieve this, we determine the number of instances based on the pair of window size and overlap that results in the minimum number of instances for the dataset. In our case, we selected a window size of 900 samples with 0\% overlap, which yielded a minimum of 1661 instances. Additionally, we ensure that the activity distribution remains consistent with the original dataset. This approach allows for a fair and standardized evaluation of different hyperparameter configurations.

\textbf{Downsampling the Dataset:}
Besides the analysed hyperparameter's, we also study the impact of the sampling frequency. For that, we conduct experiments with the following sampling frequencies: 50, 25, 12.5, 5, and 1 Hz. Since the original sampling frequency of the PAMAP2 dataset is 100 Hz, we reduce the sampling frequency of the dataset by downsampling via removing samples until we achieve the intended sampling frequency (e.g., for a frequency of 50 Hz, we remove 1/2 of the samples).

\textbf{Sampling Frequency:} \label{freq_desc}
We conduct two experiments to evaluate the effect of the sampling frequency on the accuracy, response time, and energy consumption. 
In the first experiment, we varied the sampling frequency of the train set and tested it against all considered frequencies. This allowed us to evaluate the effect of sampling frequency on system accuracy. User 9 was excluded from this experiment due to a limited number of samples. For the second experiment, we set the training set frequency at 100 Hz and tested with frequencies of 25 Hz and 1 Hz. We used the 702 configurations obtained from exploration. With this experiment, we intend to study the impact of the sampling frequency in the Pareto-Front and then compare it with the Pareto-Front obtained when the frequency is 100 Hz. When adjusting the sampling frequency of the dataset, we ensure that each sliding window contains the same number of seconds of data regardless of the frequency. This adaptation of the sliding window size is necessary to maintain consistency. However, specific constraints related to window size and overlap arise when dealing with a sampling frequency of 1 Hz. Consequently, only 189 configurations are usable for this particular frequency due to these constraints.

\section{Results} \label{results}
This section discusses the results obtained from the experiments and compares the main conclusions with the Related Work. 
\subsection{Hyperparameters}
After analyzing the results of all users, we conclude that the same configurations have different behaviors depending on the User under test, resulting in different Pareto-Fronts for each User. Table~\ref{tab:sum_all_users} summarizes the variation in the accuracy, response time, and energy consumption of the Pareto-Front of each User. 

\begin{table}[ht]
\centering
\caption{Pareto-Front values (PP) for accuracy (Acc), response time (RT), and energy consumption (EC) for all users.}
\label{tab:sum_all_users}
\begin{tabular}{ccccc}
\toprule
User & Acc (\%) & EC (mJ) & RT (ms) & PP \\ \midrule
1 & [33.05 ; 81.97] & [10.15 ; 121.12] & [5.70 ; 62.92] & 57 \\ 
2 & [26.89 ; 95.15] & [9.27 ; 116.58]  & [5.11 ; 59.64] & 64 \\
3 & [20.46 ; 93.95] & [9.16 ; 123.62]  & [5.07 ; 61.84] & 93 \\
4 & [27.85 ; 94.71] & [9.08 ; 99.85]   & [4.99 ; 51.40] & 77 \\
5 & [21.48 ; 91.06] & [9.63 ; 73.22]   & [5.29 ; 38.04] & 68 \\
6 & [27.52 ; 92.80] & [8.92 ; 105.72]  & [4.93 ; 53.11] & 71 \\
7 & [34.26 ; 95.35] & [9.30 ; 130.30]  & [5.10 ; 66.97] & 51 \\
8 & [11.08 ; 86.51] & [8.56 ; 86.55]   & [4.64 ; 44.40] & 65 \\ \bottomrule
\end{tabular}
\end{table}

Table~\ref{tab:sum_all_users}, shows that the highest accuracy measured in this experiment is 95.35\% and for User 7 using the configuration 305 (window size = 900, overlap = 0\%, distance = Manhattan, \textit{k}= 9). However, this configuration also results in the highest energy consumption (130.30 mJ) and response time (66.97 ms), considering all the Pareto-Front of all users under analysis. On the other hand, for User 1 and configuration 52 (window size = 850, overlap = 0\%, distance = Manhattan, \textit{k}= 10), the accuracy does not exceed 82\%. This User is the one that obtained the lowest maximum accuracy value of all the users studied. Despite User 1 having registered the lowest maximum accuracy, it is the third with the highest energy consumption (121.12 mJ) and the second with the highest response time (62.92 ms). There are Users with higher accuracies than User 1 and lower energy consumption and response times. For example, User 5, with configuration 909 (window size = 900, overlap = 50\%, distance = Manhattan, \textit{k}= 10), achieved 91.06\% accuracy (more 9.09\% than User 1) with the lowest energy consumption (73.22 mJ, less 47.90 mJ than User 1) and response time (38.04 ms, less 24.88 ms than User 1). Except for Users 1 and 8, the remaining users achieve a maximum accuracy above 90\%. Table~\ref{tab:sum_all_users}, shows that each User has different variations in accuracy, response time, energy consumption, and the configurations that belong to the Pareto-Front (answer to RQ 1).

\begin{figure}[ht]
  \centering
  \includegraphics[width=1\linewidth]{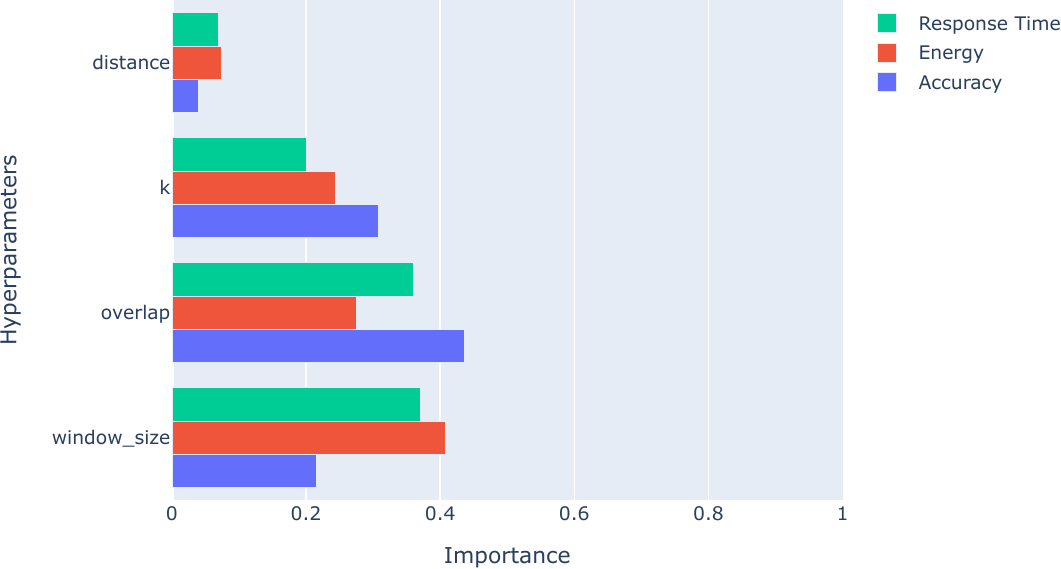}
  \caption{Hyperparameters Importance for the Pareto-front configurations.}
  \Description{Hyperparameters Importance for the Pareto-front configurations using the fANOVA algorithm for User 5.}
  \label{fig:fanova_pareto}
\end{figure}

The hyperparameters significantly impact the system's performance. However, there is a significant variation in the values for the hyperparameters. This makes it difficult to conclude which hyperparameters impact the most. To calculate the importance of each of the hyperparameters in the system, we use the fANOVA algorithm~\cite{fanova}. In this experience, we study the hyperparameters' impact on one of the 9 Users. We apply the fANOVA algorithm to the configurations of User 5. Figure~\ref{fig:fanova_pareto} shows the importance of the hyperparameters when considering the Pareto-Front configurations.

Figure~\ref{fig:fanova_pareto}, shows that window size and overlap have the most impact on the system, with the distance function being the hyperparameter with the most negligible impact. The most important factor for accuracy is the overlap. The value of \textit{k}and window size also has some importance. The importance of the distance function for accuracy is much lower. The window size and overlap are the most important energy consumption and response time hyperparameters. The results show that the window size, overlap, distance function, and \textit{k}have different levels of impact on the different performance metrics of the model and this should be considered when optimizing the model for a specific task. Depending on the desired performance metric, the relative importance of these hyperparameters may vary. For example, if energy consumption is the most important metric, then optimizing the window size and overlap may lead to the most considerable improvement in energy consumption. In conclusion, the window size and overlap are consistently important for all three performance metrics. 


Until now, we were only focusing on the global performance of the model with different configurations, but it is also important to understand the influence of the configurations at an activity level. For that, we calculate the F1-Score of the 12 activities for User 5. To better analyze the impact at an activity level, we consider the scenario where the system's accuracy must be higher than 75\%. Table~\ref{tab:f1score_per_activity} shows the F1-Score of each activity for each chosen configuration.

\begin{table*}[ht]
\centering
\caption{F1-Score of each activity for each configuration. (A1 - ascending stairs; A2 - cycling; A3 - descending stairs; A4 - ironing; A5 - lying; A6 - nordic walking; A7 - rope jumping; A8 - running; A9 - sitting; A10 - standing; A11 - vacuum cleaning; A12 - walking; WS - Window Size; Ov - Overlap; EC - Energy Consumption; RT - Response Time)}
\label{tab:f1score_per_activity}
\begin{tabular}{ccccccccccccccccccc}
\toprule
Config & A1 & A2 & A3 & A4 & A5 & A6 & A7 & A8 & A9 & A10 & A11 & A12 & AVG & WS & Ov & \textit{k}& EC & RT \\ \midrule
818 & 0.87 & 0.92 & 0.69 & 0.88 & 0.95 & 0.86 & 0.91 & 0.95 & 0.30 & 0.56 & 0.85 & 0.89 & 0.80 & 500 & 90 & 9 & 22.89 & 12.99 \\
909 & \textbf{0.90} & 0.91 & \textbf{0.88} & \textbf{0.89} & \textbf{0.98} & 0.95 & \textbf{0.97} & 0.96 & \textbf{0.89} & \textbf{0.76} & \textbf{0.88} & \textbf{0.96} & 0.91 & 900 & 50 & 10 & 73.22 & 38.04 \\
537 & 0.86 & 0.92 & 0.70 & 0.88 & 0.95 & 0.87 & 0.92 & 0.96 & 0.32 & 0.56 & 0.84 & 0.89 & 0.81 & 500 & 90 & 10 & 22.77 & 13.07 \\
893 & 0.89 & \textbf{0.94} & 0.81 & 0.84 & 0.96 & \textbf{0.96} & 0.96 & \textbf{0.97} & 0.80 & 0.67 & 0.84 & 0.93 & 0.88 & 500 & 70 & 10 & 33.53 & 18.02 \\
\bottomrule
\end{tabular}
\end{table*}

As we can see from Table~\ref{tab:f1score_per_activity}, there is a variation in the F1-Score of each activity for different configurations. There are activities where this variation is more significant. The activities running (A8), lying (A5), and cycling (A2) are the ones where the variations are smaller, 3\%, 4\%, and 6\%, respectively. The classification capability of the model for these activities is not significantly affected by the hyperparameters. These activities also have higher values for F1-Score, indicating that the model performs better in recognizing these activities. On the other hand, the activities of sitting (A9), descending stairs (A3), standing (A10), and Nordic walking (A6) are the activities where there is a significant variation in the F1-Score, with variations of 59\%, 34\%, 21\%, and 17\%, respectively. The results show that these activities strongly depend on the hyperparameter's values and the model has a lower performance in recognizing these activities. 

For the activity with higher variation in the values of F1-Score, sitting (A9), the minimum F1-Score is 0.3 (configuration 818), and the maximum is 0.89 (configuration 909). Based on the results, the value of the hyperparameters strongly influences the model's performance for this activity. Analysing Table~\ref{tab:f1score_per_activity}, we can see that the only difference in configurations 537 and 893 is the overlap, with 90\% and 70\%, respectively. This difference in the overlap allows an improvement of 0.48 in the F1-Score of the algorithm for the activity sitting (A9). 
These results show that small changes in the hyperparameters can significantly impact at an activity level.

We already concluded that the hyperparameters could impact the model's overall performance and computational cost of the HAR system. However, these results also show that the hyperparameters can significantly impact the model's performance at an activity level. For example, configuration 818 allows a global F1-Score of 0.80, but at an activity level, the performance of the model for activities sitting (A9), standing (A10), and descending stairs (A3) is significantly worse than the remaining activities. On the other hand, some activities are not affected by the variations of the hyperparameter's values, in the case of activities like running (A8), lying (A5), and cycling (A2). These results also provided important information regarding energy consumption. In a scenario where the user of the system is performing a specific activity for an extended time (for example, A8 - running), the system can change to a configuration (for example, from 909 to 537) that maintains the recognition performance of the model for the activity (F1-Score of 0.96) but requires less energy (73.22 mJ to 22.77 mJ), even though the global performance of the system decreases (0.91 to 0.81). These results reinforced the importance of adaptability in HAR systems because although configuration provides a global high performance for the system, some activities may be negatively affected by the configuration (answer to RQ 2).

\subsection{Sampling Frequency}
Table~\ref{freqs_combo} summarizes the combinations of the sampling frequency for the train and test sets that allow us to obtain the maximum and minimum accuracy.

\begin{table}[ht]
\begin{center}
\caption{Combinations of sampling frequency for the training and test sets that obtained the maximum and minimum accuracy for each User and respective accuracy (Acc) values.}
\label{freqs_combo}
\begin{tabular}{ccccccc}
\multirow{2}{*}{User} & \multicolumn{3}{c}{Max (Hz)} & \multicolumn{3}{c}{Min (Hz)} \\ \cline{2-7} 
 & Train & Test & Acc (\%) & Train & Test & Acc (\%) \\ \hline
1 & 5 & 5 & 80.50 & 1 & 1 & 71.48 \\ \hline
\multirow{2}{*}{2} & 50 & 5 & 93.86 & 1 & 5 & \multirow{2}{*}{85.62} \\
 & - & - & - & 1 & 25 &  \\ \hline
\multirow{3}{*}{3} & 100 & 12.5 & \multirow{3}{*}{93.23} & 1 & 1 & 80.83 \\
 & 25 & 12.5 &  & - & - & - \\
 & 5 & 12.5 &  & - & - & - \\ \hline
4 & 25 & 12.5 & 94.14 & 1 & 1 & 81.71 \\ \hline
\multirow{2}{*}{5} & 100 & 50 & \multirow{2}{*}{86.09} & 5 & 1 & 80.46 \\
 & 25 & 50 &  & - & - & - \\ \hline
\multirow{2}{*}{6} & 100 & 12.5 & \multirow{2}{*}{92.39} & 25 & 1 & \multirow{2}{*}{82.67} \\
 & 50 & 12.5 &  & 5 & 1 &  \\ \hline
7 & 5 & 5 & 95.74 & 12.5 & 1 & 91.47 \\ \hline
8 & 25 & 100 & 85.22 & 1 & 1 & 68.38 \\ \hline
\end{tabular}
\end{center}
\end{table}

The results show that changing the sampling frequencies of the training set and of the test set significantly impacted the model's accuracy. Overall, all users show a decreased accuracy when the sampling frequency is reduced. However, the decrease in the sampling frequency impacted all users differently. There are users for whom this impact is more significant than others. The biggest drop in the accuracy is shown for User 8, with a decrease in accuracy of about 17\%. In this case, the sampling frequency of the training set used to train the model is 1 Hz, and the sampling frequency of the test set is also 1 Hz. There are also other significant drops in the accuracy for Users 3 and 4, where we obtained a decrease of 12.40\% and 12.43\%, respectively. On the other hand, User 7 reports a slight reduction in accuracy, dropping 4.27\%, and User 5 shows a decrease of 5.63\% (answer to RQ 3). 

The combination of the sampling frequency for the train and test sets that resulted in the lowest accuracy obtained is when both sets have a sampling frequency of 1 Hz (see Table~\ref{freqs_combo}). This combination obtains lower accuracy for 50\% of the users (Users 1, 3, 4, and 8). Although we report a decrease in the accuracy for all users when changing the sampling frequency, except for Users 1 and 8, all the remaining users obtain minimum accuracy greater than 80\%. Regarding Users 1 and 8, the minimum accuracy obtained is 71.48\% and 63.38\%, respectively. Considering the scenario where the acceptable accuracy for a HAR system is greater than 75\%, the sampling frequency of 1 Hz for the train and test sets for Users 1 and 8 are not sufficient to be used in the system.

Although for Users 1 and 7 the highest accuracy achieved is with the model trained and tested with a sampling frequency of 5 Hz, on average, the highest accuracy values are obtained for sampling frequencies greater or equal to 12.5 Hz for the train and test sets. These results are observed for all users. Overall, the accuracy benefits from sampling frequencies greater or equal to 12.5 Hz, while sampling frequencies equal to or below 5 Hz show significant decreases in accuracy (answer to RQ 5). 

\begin{figure}[ht]
  \centering
  \includegraphics[width=1\linewidth]{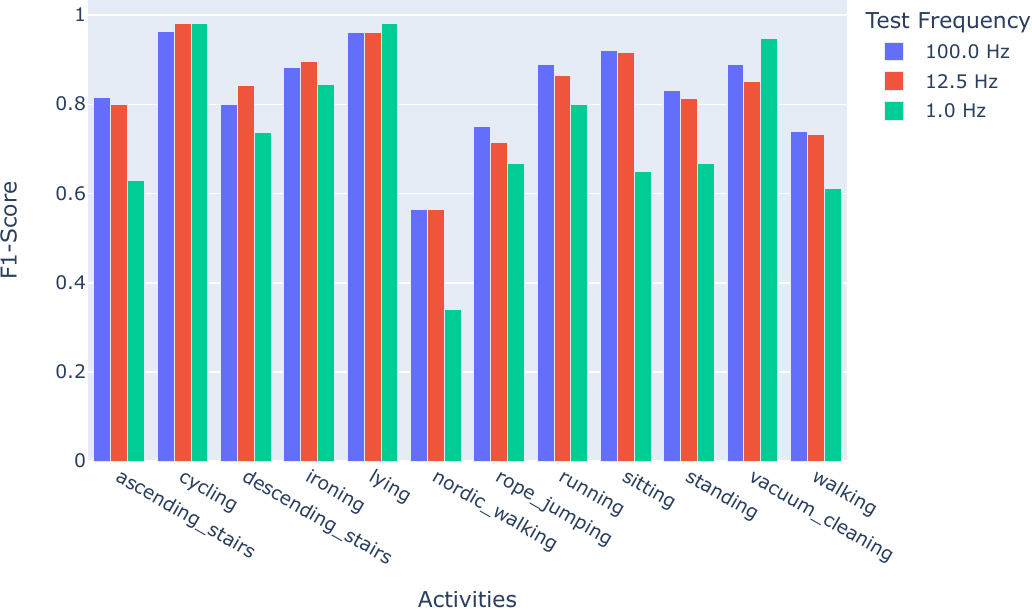}
  \caption{F1-Score per activity for different values of the sampling frequency of the test set, maintaining the sampling frequency of the trained model as 100 Hz (User 8).}
  \Description{Results of the F1-Score metric when changing the sampling frequency of the test set, while maintaining the sampling frequency of the train set, for User 8.}
  \label{fig:different_test_freq_user8}
\end{figure}

Understanding how the sampling frequency affects the algorithm's performance at the activity level is also important. Different sampling frequencies may have other effects depending on the activity. For that, we show the F1-Score for each activity for different sampling frequencies for User 8, because it is the User with whom the sampling frequency has more impact, where there is a variation of 17\% between the highest and lower accuracy. Figure~\ref{fig:different_test_freq_user8} shows the effect of the sampling frequency on the activities of PAMAP2 when using the same sampling frequency of the train set (100 Hz) and vary the sampling frequency of the test set for 100 Hz, 12.5 Hz, and 1 Hz. 

Figure~\ref{fig:different_test_freq_user8} shows that the sampling frequency of the test set has different effects on the activities of the PAMAP2 dataset. Varying the sampling frequency of the test set does not significantly affect the model's performance for cycling, lying, and vacuum cleaning activities. There is a decreased performance for most activities when the sampling frequency is reduced. This decrease is more significant for Nordic walking, walking, standing, sitting, and ascending stairs. For these activities, the F1-Score decreased with the sampling frequency reduction. The lowest values of the F1-Score are obtained for a sampling frequency of 1 Hz. "Sitting" and "Nordic walking" activities show a more significant decrease in performance when the sampling frequency changes from 100 Hz to 1 Hz. However, there is an increase in F1-Score for "vacuum cleaning" when lower sampling frequencies are used for the test set.

Although these results suggest that higher sampling frequencies are generally better for improving the performance of \textit{k}NN in recognizing different activities, the optimal sampling frequency may depend on the recognized activity. For example, \textit{Nordic walking} is better recognized at 100 Hz than at 1 Hz. The results show that the choice of sampling frequency can significantly impact the performance of the \textit{k}NN model, as demonstrated by the varying performance across different activities (answer to RQ 3).

\begin{figure}[ht]
  \centering
  \includegraphics[width=1\linewidth]{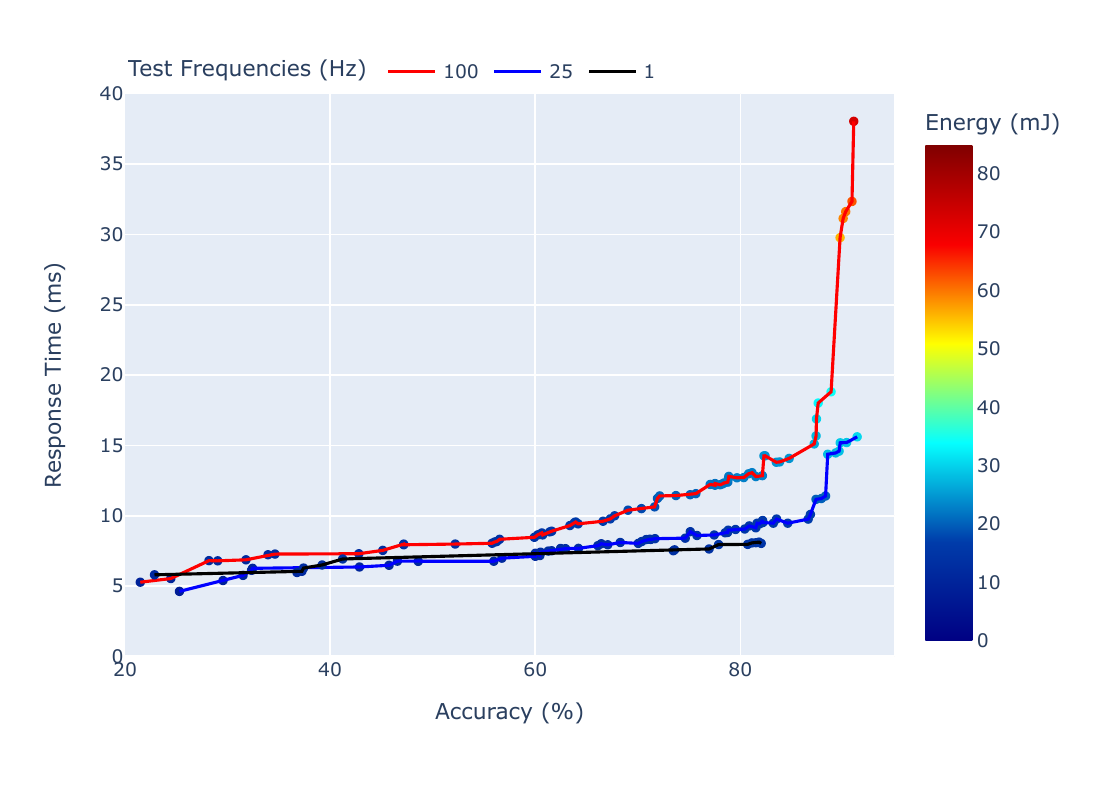}
  \caption{Pareto-frontier of User 5 for different test frequencies (100 Hz, 25 Hz, and 1 Hz).}
  \Description{Pareto-Front for the sampling frequencies of 100 Hz, 25 Hz, and 1 Hz for the test set of User 5. The sampling frequency of the train set is the same for all cases (100 Hz).}
  \label{fig:pareto_freq}
\end{figure}

We also evaluate the effect of the sampling frequency on the Pareto-Front of User 5. We use a sampling frequency of 100 Hz for the train set and then test them using sampling frequencies of 100 Hz, 25 Hz, and 1 Hz. It is important to note that the window sizes of the configurations are adjusted to the sampling frequency to maintain the same number of seconds. The results of this experiment are presented in Figure~\ref{fig:pareto_freq}. We show the Pareto-Front of the original sampling frequency (red line), of the 25 Hz frequency (blue line), and of the 1 Hz frequency (black line). Table~\ref{tab:freq_var} presents the lower and upper extreme points of the Pareto-Front for the sampling frequencies of 100 Hz, 25 Hz, and 1 Hz. We also present the variation in response time, energy consumption, and accuracy. 

Figure~\ref{fig:pareto_freq} shows that the Pareto-Fronts are different for each of the frequencies under study. The configurations that constitute the Pareto-Fronts are different for each sampling frequency, and the number of configurations present in the frontiers is also different. The sampling frequency of 100 Hz has 68 configurations, 25 Hz has 60 configurations, and 1 Hz has only 15 configurations. Focusing only on the sampling frequency of 100 Hz, we observe a significant increase in the energy consumption and response time for accuracies greater or equal to 85\%, with a slight increase in the accuracy. However, this behavior is absent for 25 Hz and 1 Hz sampling frequencies. When the sampling frequency is reduced, the energy consumption and response time variation become smaller (see Table~\ref{tab:freq_var}). The Pareto-Front for the sampling frequency of 1 Hz is where the response time and energy consumption have the lowest variation. For this Pareto-Front, there is a variation of 3.62 mJ in energy consumption and 2.23 ms in response time between the lower and upper extremes of the frontier. At the same time, the accuracy registered an increase of about 59.16\%. 

\begin{table}[ht]
\centering
\caption{Variation of the accuracy (Acc), response time (RT), and energy consumption (EC) for the sampling frequencies of 100 Hz, 25 Hz, and 1 Hz. (Freq - Frequency (Hz); LP - Lower Point; UP - Upper Point)}
\label{tab:freq_var}
\begin{tabular}{cccccc}
\toprule
Freq & LP & UP & Acc (\%) & EC (mJ) & RT (ms)  \\ \midrule
100 & 842 & 909 & [21.48; 91.06] & [9.63; 73.22] & [5.29; 38.40]  \\ 
25 & 458 & 944 & [25.30; 91.39] & [8.18; 30.63]  & [4.63; 15.62]  \\
1 & 226 & 682 & [22.87; 82.03] & [11.19; 14.81] & [5.81; 8.04]  \\ \bottomrule
\end{tabular}
\end{table}

Comparing the Pareto-Front for the sampling frequencies of 100 Hz and 25 and focusing only on the configuration that allows us to obtain the maximum accuracy (909 for 100 Hz and 944 for 25 Hz, see Table~\ref{tab:freq_var}), we can observe a significant reduction in energy consumption and response time. We measured a decrease of 58\% in energy consumption and 59\% in response time. At the same time, the accuracy does not suffer significant alterations. Considering now the Pareto-Front for the sampling frequencies of 100 Hz and 1 Hz and focusing again only on the configuration that allows us to obtain the maximum accuracy (909 for 100 Hz and 682 for 1 Hz, see Table~\ref{tab:freq_var}), it is possible to observe a higher reduction on the energy consumption and response time. We observe a decrease of 80\% in energy consumption and 79\% in response time. However, the sampling frequency of 1 Hz also results in a significant reduction in the accuracy of about 11\%. Considering the scenario where we assume that an acceptable accuracy is above 75\%, both 25 Hz and 1 Hz sampling frequencies have configurations above the threshold, allowing the use of these frequencies in this hypothetical system (answer to RQ 4). 
 
From this experiment, smaller sampling frequencies significantly reduce the system's energy consumption and response time without losing significant accuracy (keeping the same seconds per window). We note that the results of this experiment and its conclusions are for User 5, we cannot generalize these results to the other Users since each User data has its behavior and characteristics that can have different effects when the sampling frequency is tested.

\subsection{Analysis with Respect to Related Work}
During this work, we conducted several experiments to evaluate the impact of some parameters (window size, overlap, the value of \textit{k}, distance function, and sampling frequency) on the performance of a HAR system (accuracy, F1-Score, energy consumption, and response time). As our experimental results and conclusions adrresses the PAMAP2 dataset and the \textit{k}NN algorithm, we compare them to the findings of the works identified in Section~\ref{related}. Table~\ref{tab:related_work} resumes the Related Work, to be compared with our study. We analyze each hyperparameter individually to evaluate the impact on the system. To do this, we fixed the values of the remaining hyperparameter and select the configurations where only the hyperparameter under study changes. Table~\ref{tab:fixed_vaues} shows the values for the fixed hyperparameters and those used for the hyperparameter under study. 

\begin{table}[ht]
\centering
\caption{Hyperparameter values used to evaluate their individual influence on the HAR system. }
\label{tab:fixed_vaues}
\begin{tabular}{c|ccc}
\toprule
 & Window Size (WS) & Overlap (\%) & \textit{k}(\textit{k}NN) \\ 
\midrule
WS & \begin{tabular}[c]{@{}c@{}}100, 150, 250, 300, 350, \\500, 550, 600, 650, 750, \\800, 850, 900\end{tabular} & 250 & 250 \\ \hline
Overlap & 50 & \begin{tabular}[c]{@{}c@{}}0, 10, 30, 40, \\ 60, 70, 80, 90\end{tabular} & 80 \\ \hline
k & 10 & 9 & \begin{tabular}[c]{@{}c@{}}1, 2, 3, 5, \\ 6, 9, 10\end{tabular} \\ 
\bottomrule
\end{tabular}
\end{table}

As we already concluded, each User has its behavior when using the same configurations. To compare our results with the Related Work, we first need to know the influence of each hyperparameter individually for each User. We calculate the Pearson Correlation between the hyperparameters studied and the performance metrics used to evaluate the system. Table~\ref{tab:correlations} shows the values of the correlations between the hyperparameters and performance metrics for each User. 

\begin{table}[ht]
\tiny
\centering
\caption{Pearson Correlation between each hyperparameter and each performance metric (Acc - Accuracy; EC - Energy Consumption; RT - Response Time).}
\label{tab:correlations}
\begin{tabular}{c|ccc|ccc|ccc|}
\multirow{2}{*}{User} & \multicolumn{3}{c|}{Window Size} & \multicolumn{3}{c|}{Overlap} & \multicolumn{3}{c|}{k (\textit{k}NN)} \\ \cline{2-10} 
 & Acc & EC & RT & ACC & EC & RT & ACC & EC & RT \\ \hline
1 & 0.5456 & 0.9994 & 0.9992 & -0.5055 & -0.9991 & -0.9996 & 0.9830 & 0.9407 & 0.8265 \\
2 & 0.9454 & 0.9992 & 0.9993 & -0.9397 & -0.9983 & -0.9953 & 0.9238 & 0.8856 & 0.9670 \\
3 & 0.8939 & 0.9993 & 0.9987 & -0.7819 & -0.9997 & -0.9998 & 0.8717 & 0.8587 & 0.7764 \\
4 & 0.9100 & 0.9993 & 0.9990 & -0.9696 & -0.9993 & -0.9997 & 0.9481 & 0.8184 & 0.8769 \\
5 & 0.8618 & 0.9996 & 0.9992 & -0.7672 & -0.9985 & -0.9975 & 0.9777 & 0.9455 & 0.9467 \\
6 & 0.7969 & 0.9992 & 0.9992 & -0.9457 & -0.9993 & -0.9987 & 0.8505 & 0.8784 & 0.8743 \\
7 & 0.3993 & 0.9992 & 0.9993 & -0.8288 & -0.9997 & -0.9994 & 0.6678 & 0.8965 & 0.6645 \\
8 & 0.8437 & 0.9984 & 0.9993 & -0.2874 & -0.9997 & -0.9996 & 0.9932 & 0.9404 & 0.8393 \\ \hline
\end{tabular}
\end{table}

Window size positively correlates with accuracy, energy consumption, and response time. This indicates that these performance metrics increase when the window size also increases. This correlation is stronger for energy consumption and response time. Regarding accuracy, the correlation is stronger for some users. For example, for User 7, the correlation between window size and accuracy is weaker than for the remaining Users. For this specific case, the accuracy increases (from 71\% to 89\%) until a window size of 500, and after that, it starts decreasing (from 89\% to 82\%). Overall, larger window sizes, at some extent, lead to higher accuracy. 

The overlap increase leads to decreased accuracy, energy consumption, and response time, per inference. However, the increase in the overlap leads to more inferences and globally to an increase in the energy consumption. The negative correlation between overlap and the performance metrics confirms this. Like window size, the correlation is stronger for energy consumption and response time. For all Users, higher overlapping decreases energy consumption and response time. As for accuracy, some users are less affected by the overlap, such as Users 1 and 8. For these Users, the influence of overlap in the accuracy differs significantly from the remaining Users. For example, the highest accuracy is achieved for an overlap of 60\%. In contrast, for the remaining users, the maximum accuracy is achieved for an overlap inferior to 50\% and superior or equal to 10\%. The overlap values between 10\% and 50\% allow better accuracy. For overlaps higher than 50\%, the accuracy starts to decrease significantly. 

The value of \textit{k}is the hyperparameter with a weaker influence on the performance metrics. Changes in the \textit{k}lead to small accuracy, energy consumption, and response time changes. For most users, these changes are inferior to 5\% for accuracy, 2 mJ for energy consumption, and 2 ms for response time. 
However, overall, higher values of \textit{k}lead to higher accuracies, energy consumption, and response time. 

Mohsen et al.~\cite{Mohsen2022} and Liu et al.~\cite{Liu2021} study the influence of the value of \textit{k}on a \textit{k}NN-base HAR system. Mohsen et al.~\cite{Mohsen2022} evaluate their studies on the UCI-HAR dataset and conclude that the accuracy increases with the increase of the value of k. The best accuracy is obtained using values of \textit{k}between 8 and 20. Increasing the value of \textit{k}above 8 does not yield any significant improvement. On the other hand, Liu et al.~\cite{Liu2021} evaluate the effect of different values of \textit{k}, between 3 and 9, on the accuracy using the HAPT~\cite{hapt_dataset} and Smartphone~\cite{Dua2019} dataset. They conclude that increasing the value of \textit{k}increases the accuracy of the models. However, beyond a \textit{k}value of 6, the accuracy decreases with the increase of k. Although our results on the effect of the value of \textit{k}on the HAR system are aligned with the related work, we do not observe significant increases in accuracy. The accuracy increase with the increase of the value of \textit{k}, is inferior to 5\%. Also, we do not observe a decrease in the accuracy for \textit{k}values beyond 6, such as Liu et al.~\cite{Liu2021}. As expected, we also observe an increase in energy consumption and response time when increasing the value of k. However, we cannot compare it with the related work since none of the authors explored the effect of the value of \textit{k}in these performance metrics. 

Another hyperparameter explored in the work presented in the Related Work is overlap. Garcia et al.~\cite{kemilly} study the influence of the overlap in accuracy, energy consumption, and execution time for the PAMAP2 dataset. They conclude that the overlap shows more fluctuations in terms of accuracy. The best values are and as in our case, between 10\% and 50\%. Increasing the overlap factor also increases the energy spent and execution time. Our results are similar to the author's, and larger overlappings lead to higher energy consumption and execution time. However, when considering per inference instead globally, we concluded that higher overlapping leads to decreased energy consumption and response time.

The window size is one of the most studied hyperparameters in HAR research. Banos et al.~\cite{Banos2014} study the influence of window size, on the REALDISP dataset, between 0.25 s and 7 s. They conclude that windows with a size of 2 seconds achieved the best results and larger window sizes do not always lead to improved performance. Garcia et al.~\cite{kemilly} explored window sizes ranging from 100 to 1000 samples on the PAMAP2 dataset, considering accuracy, energy consumption, and execution time. They found that smaller window sizes resulted in lower accuracy, while larger sizes improved accuracy. Additionally, smaller windows consumed less energy and had shorter execution times compared to larger windows. Wang et al.~\cite{wang_2018} utilized a private dataset and investigated window sizes between 0.5 and 7 seconds. They observed a significant enhancement in classification performance with larger window sizes. However, larger window sizes also introduced increased recognition latency. Niazi et al.\cite{Niazi2017} evaluated window sizes of 1, 2, 3, 5, and 10 seconds for a private dataset. They found that better results were obtained with larger window sizes. Comparing our findings with these studies, we confirm that larger window sizes generally lead to improved recognition performance, as observed by Garcia et al.\cite{kemilly}, Wang et al.\cite{wang_2018}, and Niazi et al.\cite{Niazi2017}. Similar to Wang et al.\cite{wang_2018} and Garcia et al.\cite{kemilly}, we show that larger window sizes result in longer response times and higher energy requirements. However, our results differ from Banos et al.~\cite{Banos2014}, as we found that larger windows can increase the model's accuracy.

Although the sampling frequency is not a hyperparameter, it can significantly influence the system's performance. Niazi et al.~\cite{Niazi2017} use 5, 10, 20, 25, 50, and 100 Hz for sampling frequency. The best results are spread around high sampling rates. Zheng et al.~\cite{Zheng2017} use 1, 5, 10, and 50 Hz sampling frequencies and evaluate their impact on the accuracy and energy consumption. The results show that the accuracy has only improved slightly with the sampling rate increase from 1 to 50 Hz. In terms of energy, there is an increase in energy consumption with increased sampling frequency. Santoyo-Ramón et al.~\cite{Ramon2022} show that the system performance only degrades when the sampling rate is set to 10–15 Hz. The detection ratio improves when the initial rate is reduced to values between 20 and 40 Hz. The conclusions achieved by Santoyo-Ramón et al.~\cite{Ramon2022} is the most similar to ours. Regarding the model's performance, we show no significant difference in sampling frequencies between 12.5 and 100 Hz. However, specifically, for a sampling frequency of 1 Hz, we observe a significant degradation in the model's performance. For the computational cost, smaller sampling frequencies lead to lower energy consumption and response time. Zheng et al.~\cite{Zheng2017} also confirm that higher sampling frequencies translate into higher energy consumption. They also achieve the lowest accuracy for a sampling frequency of 1 Hz. Our conclusions differ from Niazi et al.~\cite{Niazi2017} as they observed better results for higher sampling frequencies (in the case of 50 Hz). However, we verify that for some cases, sampling frequencies of 12.5 Hz achieve better results than the other higher frequencies. 

Overall, our result is aligned with most of the work identified in the Related Work, even though different datasets and inference algorithms are being used. Our conclusion is stronger in similarity to the conclusions of Garcia et al.~\cite{kemilly} since both use the same dataset and experimental setup. However, there are some exceptions. For example, Niazi et al.~\cite{Niazi2017} show that the best model performance is achieved for a maximum sampling frequency of 50 Hz. Our results show that lower sampling frequencies perform better than bigger ones in some situations. The conclusion for the sampling frequency is more influenced by the datasets used since a large variety of devices were used to collect the data. Most of the studies presented in the Related Work evaluate the influence on the model's performance. Only some exceptions (Garcia et al.~\cite{kemilly} and Zheng et al.~\cite{Zheng2017}) have studied the influence on energy consumption and execution time. When targeting the implementation of a HAR system on a device to use in real life, it is essential to consider, in addition to the model's performance, its impact on the device in terms of energy consumption and response time.

\section{Conclusion} \label{conclusion}

Human activity recognition (HAR) is a research area that has been increasingly addressed due to the widespread of mobile devices such as smartphones. These devices have enough processing power and memory to implement HAR systems. The more sensors embedded in these devices also allow a more accurate HAR system. However, the use of more sensors can increase energy consumption, which decreases the device's battery life. 

HAR systems must be able to adapt to the device constraints, the user, and the environment to be more efficient. For example, when the device is running out of battery, the system should be able to adjust its behavior to reduce energy consumption and, in this way, increase the battery life of the device. In addition, people have different walking and motion patterns, and these patterns for the same person may change during a lifetime, caused by aging, diseases, or accidents. Not addressing this problem can result in a loss of accuracy.  

One way to change the system's behavior is by changing its hyperparameters. 
To understand how the hyperparameters can influence the system, in this paper, we studied the impact of a selected set of hyperparameters, namely window size, overlap, the value of \textit{k}, and distance function, and also sampling frequency, on a \textit{k}NN-based HAR system in terms of accuracy, response time, and energy consumption, using the PAMAP2 dataset.


Based on the obtained results, we conclude that hyperparameters significantly impact the system's accuracy, response time, and energy consumption and that the same configurations have different implications for different users. Each user had a different number and configurations belonging to the Pareto-Front, and the range values of the performance metrics were also different. For example, User 1 requires more energy to achieve an accuracy of 82\% than most users to achieve an accuracy higher than 90\%. In another example, Users 1 and 8 did not achieve an accuracy higher than 82\% and 87\%, respectively, while all other users achieved an accuracy greater than 90\%. These results reinforce the importance of studying the hyperparameters and the necessity of a system adapting to the user. 

Overall, window size and overlap were the hyperparameters with a more significant impact on the accuracy, response time, and energy consumption. Windows with more samples require more time and energy to extract the features necessary to train and test the models. Higher overlaps imply more calls to the inference task and globally more energy consumption.

At activity levels, there is a need to consider adaptability. A configuration with high global performance for some activities may not be the most adequate for a specific activity. 
The results show the importance of considering the activities individually when developing a HAR system instead of considering only the global performance of the model. 

The sampling frequency also significantly impacts a HAR system and has different behaviors depending on the User. For example, there are Users much more affected by the reduction of the sampling frequency. Once again, this emphasizes the importance of a self-adaptive system. The sampling frequency also has more impact on some activities. Although lowering the sampling frequency leads to lower energy consumption and response time, it may impact the accuracy of the system. 

When implementing a HAR system on a device for use in real-life scenarios, it is important to evaluate both the performance of the model and the impact on the device regarding energy consumption and response time. We conclude that the hyperparameters significantly impact a HAR system and should be studied before implementing the system. These conclusions also motivate continuing this work and the necessity of a self-adaptive HAR system for the user, the device, and the target service.

In future work, we intend to continue studying hyperparameters' impact on a HAR system and consider other ML algorithms and datasets.


\clearpage



\bibliographystyle{ACM-Reference-Format}
\bibliography{sample-base}
\onecolumn
\appendix
\section{Appendix}
\subsection{Comparison with Related Work}
\begin{table*}[ht!]
\centering
\caption{Detailed information regarding the Related Work.}
\label{tab:related_work}
\begin{tabular}{cccccc}
Study & Hyperparameters & Dataset & \begin{tabular}[c]{@{}c@{}}Number of\\ Activities\end{tabular} & Metrics & \begin{tabular}[c]{@{}c@{}}Inference \\ Algorithm\end{tabular} \\ \hline
\begin{tabular}[c]{@{}c@{}}Banos et al.\\ ~\cite{Banos2014}\end{tabular} & \begin{tabular}[c]{@{}c@{}}Window Size (s) \\ (0.25 to 7)\end{tabular} & REALDISP~\cite{realdisp} & 33 & F1-Score & \begin{tabular}[c]{@{}c@{}}C4.5, \textit{k}NN,\\ Naïve Bayes,\\ Nearest \\ Centroid\end{tabular} \\ \hline
\begin{tabular}[c]{@{}c@{}}Garcia et al.\\ ~\cite{kemilly}\end{tabular} & \begin{tabular}[c]{@{}c@{}}Window Size\\ (samples)\\ (100 to 1000)\\ Overlap (0.0 to 0.9)\end{tabular} & PAMAP2~\cite{pamap,pamap2} & 12 & \begin{tabular}[c]{@{}c@{}}Accuracy,\\ Energy, \\ Execution \\ Time\end{tabular} & \begin{tabular}[c]{@{}c@{}}\textit{k}NN, VFDT,\\ Naive Bayes,\\ Ensemble\end{tabular} \\ \hline
\begin{tabular}[c]{@{}c@{}}Wang et al.\\ ~\cite{wang_2018}\end{tabular} & \begin{tabular}[c]{@{}c@{}}Window Size (s)\\ (0.5 to 7)\end{tabular} & Private & 12 & F1-Score & \begin{tabular}[c]{@{}c@{}}SVM, \textit{k}NN,\\ Decision Tree,\\ Naïve Bayes,\\ Adaboost\end{tabular} \\ \hline
\begin{tabular}[c]{@{}c@{}}Mohsen et al.\\ ~\cite{Mohsen2022}\end{tabular} & \begin{tabular}[c]{@{}c@{}}k (1 to 20)\end{tabular} & UCI-HAR~\cite{uci-har} & 6 & Accuracy & \textit{k}NN \\ \hline
\begin{tabular}[c]{@{}c@{}}Liu et al.\\ ~\cite{Liu2021}\end{tabular} & \begin{tabular}[c]{@{}c@{}}k (3 to 9)\end{tabular} & \begin{tabular}[c]{@{}c@{}}HAPT~\cite{hapt_dataset}\\ Smartphone~\cite{Dua2019}\end{tabular} & \begin{tabular}[c]{@{}c@{}}9\\ 6\end{tabular} & Accuracy & \textit{k}NN \\ \hline
\begin{tabular}[c]{@{}c@{}}Ramón et al.\\ ~\cite{Ramon2022}\end{tabular} & \begin{tabular}[c]{@{}c@{}}Sampling\\ Frequency (Hz) \\ (18 to 238)\end{tabular} & \begin{tabular}[c]{@{}c@{}}15 HAR \\ Datasets\end{tabular} & 6 to 44 & \begin{tabular}[c]{@{}c@{}}Mean \\ geometric \\ of \\ sensitivity \\ and \\ specificity\end{tabular} & CNN \\ \hline
\begin{tabular}[c]{@{}c@{}}Niazi et al.\\ ~\cite{Niazi2017}\end{tabular} & \begin{tabular}[c]{@{}c@{}}Sampling \\ Frequency (Hz)\\ (5, 10, 20, \\ 25, 50, 100)\\ Window Size (s)\\ (1, 2, 3, 5, 10)\end{tabular} & Private & 23 & \begin{tabular}[c]{@{}c@{}}Average \\ expected \\ value \end{tabular}& \begin{tabular}[c]{@{}c@{}}Random \\ Forest\end{tabular} \\ \hline
\begin{tabular}[c]{@{}c@{}}Zheng et al.\\ ~\cite{Zheng2017}\end{tabular} &  \begin{tabular}[c]{@{}c@{}}Sampling \\ Frequency (Hz)\\ (1, 5, 10, 50)\end{tabular} & Private & 6 & \begin{tabular}[c]{@{}c@{}}Accuracy\\ Energy\end{tabular} & SVM \\ \hline
This Work & \begin{tabular}[c]{@{}c@{}}Sampling \\ Frequency (Hz)\\ (1, 5, 12.5, 25,\\ 50, 100)\\ Window Size\\ (samples)\\ (50 to 900)\\ Overlap (0 to 0.9)\\ \textit{k}(1 to 10)\\ Distance (E, M, C)\end{tabular} & PAMAP2~\cite{pamap,pamap2} & 12 & \begin{tabular}[c]{@{}c@{}}Accuracy,\\ F1-Score,\\ Energy,\\ Inference\\ Time\end{tabular} & \textit{k}NN \\ \hline
\end{tabular}
\end{table*}

\subsection{Extended Version} \label{ext}


\section*{Supplementary material}

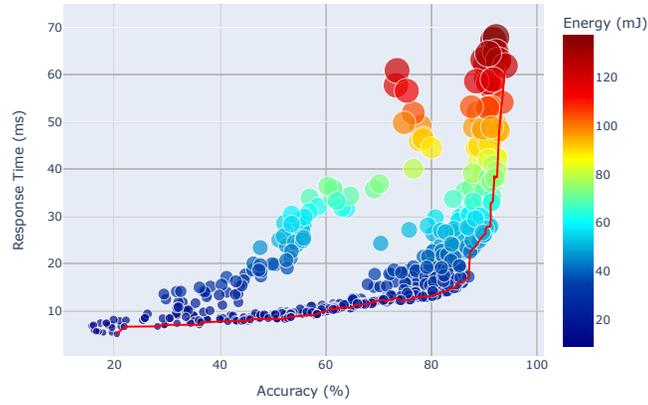

**Fig. 20.** Energy Consumption, Response Time, and Accuracy of the configurations for User 3.

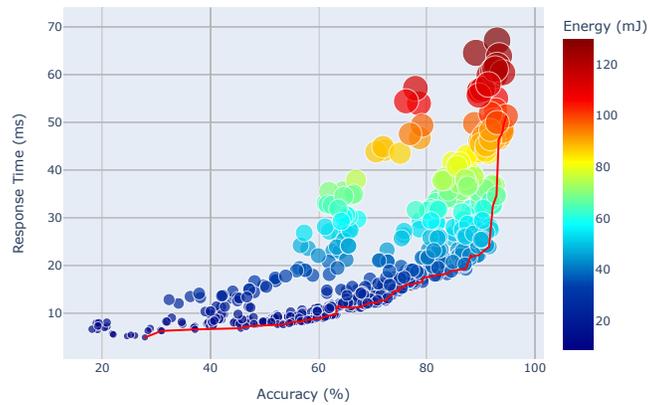

**Fig. 21.** Energy Consumption, Response Time, and Accuracy of the configurations for User 4.

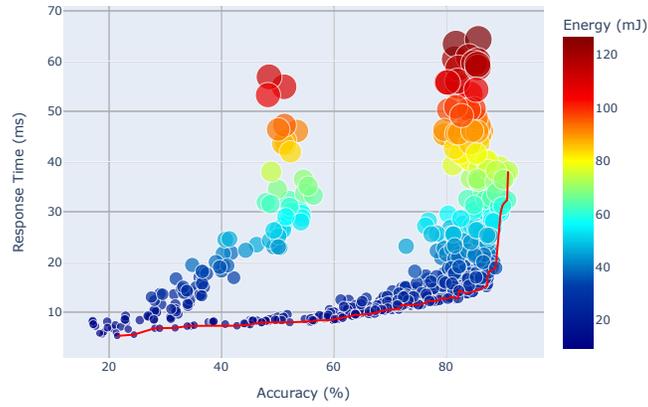

**Fig. 22.** Energy Consumption, Response Time, and Accuracy of the configurations for User 5.

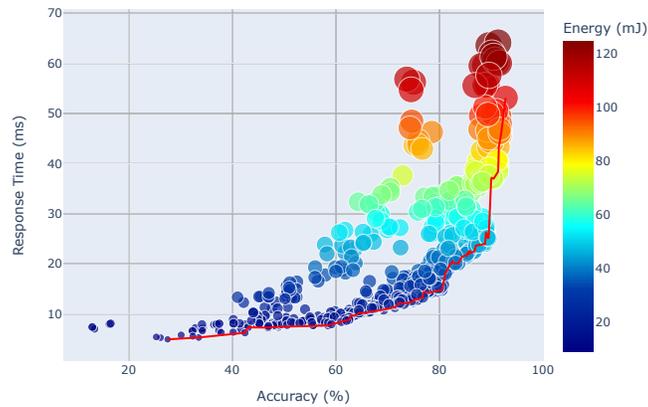

**Fig. 23.** Energy Consumption, Response Time, and Accuracy of the configurations for User 6.

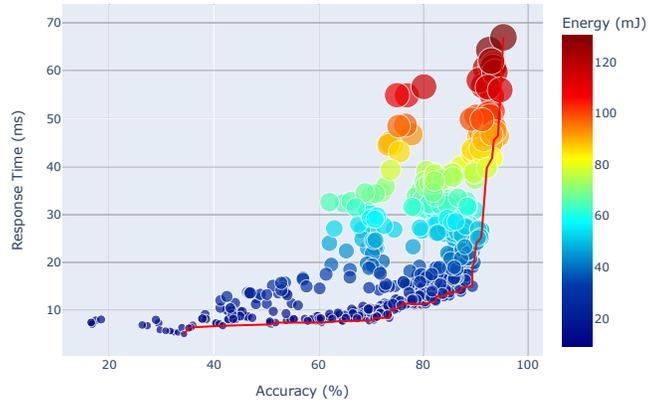

**Fig. 24.** Energy Consumption, Response Time, and Accuracy of the configurations for User 7.

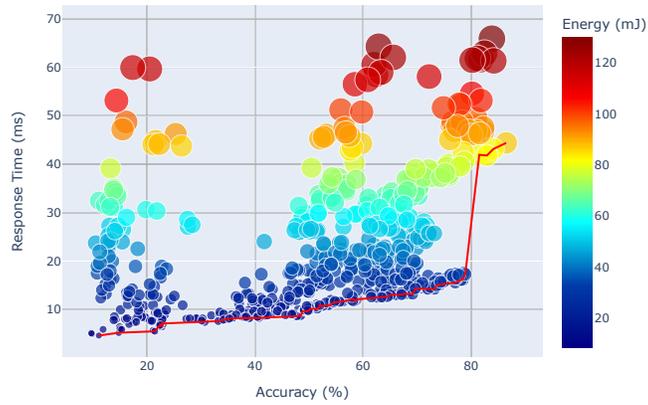

**Fig. 25.** Energy Consumption, Response Time, and Accuracy of the configurations for User 8.

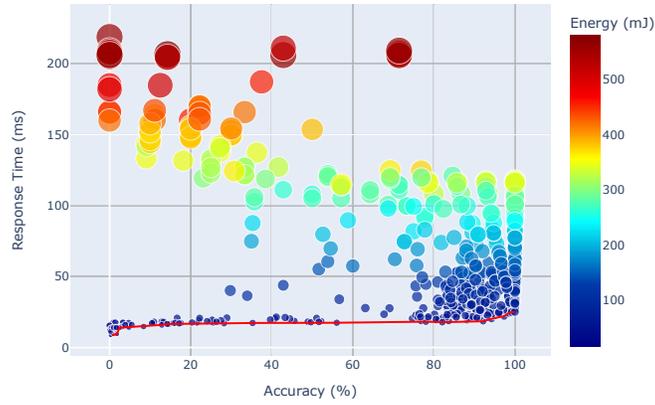

**Fig. 26.** Energy Consumption, Response Time, and Accuracy of the configurations for User 9.

## A.2   Pareto-Front configurations with an accuracy greater or equal to 75%

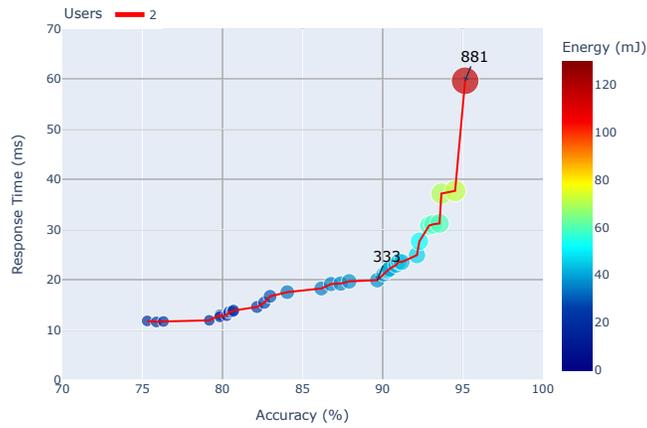

**Fig. 27.** Pareto-Front for User 2 with configurations with an accuracy greater than 75%. Some configurations highlighted (configuration ID is shown).

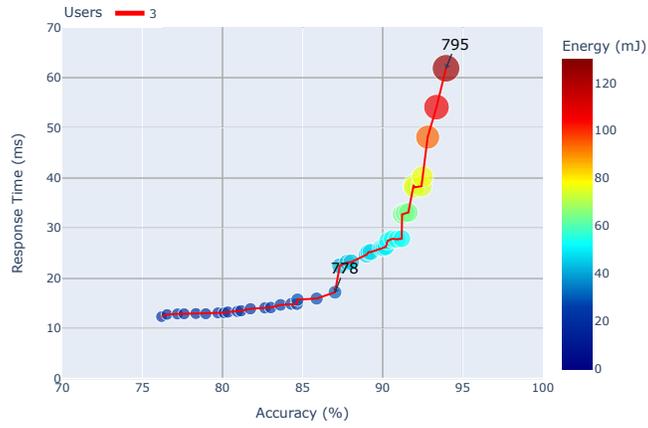

**Fig. 28.** Pareto-Front for User 3 with configurations with an accuracy greater than 75%. Some configurations highlighted (configuration ID is shown).

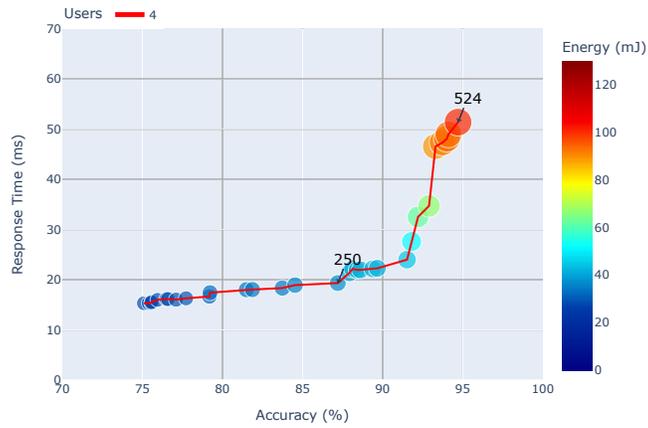

**Fig. 29.** Pareto-Front for User 4 with configurations with an accuracy greater than 75%. Some configurations highlighted (configuration ID is shown).

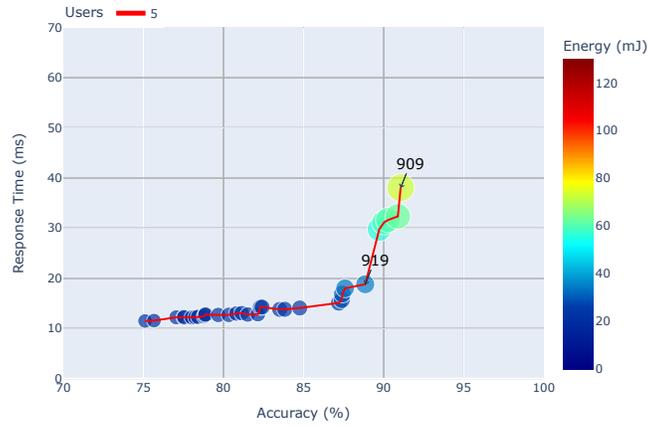

**Fig. 30.** Pareto-Front for User 5 with configurations with an accuracy greater than 75%. Some configurations highlighted (configuration ID is shown).

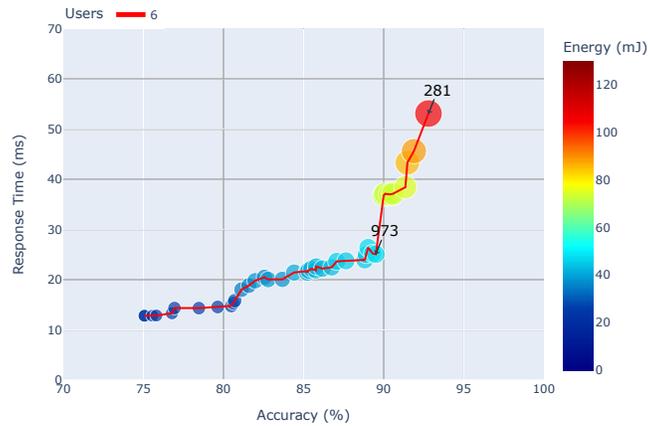

**Fig. 31.** Pareto-Front for User 6 with configurations with an accuracy greater than 75%. Some configurations highlighted (configuration ID is shown).

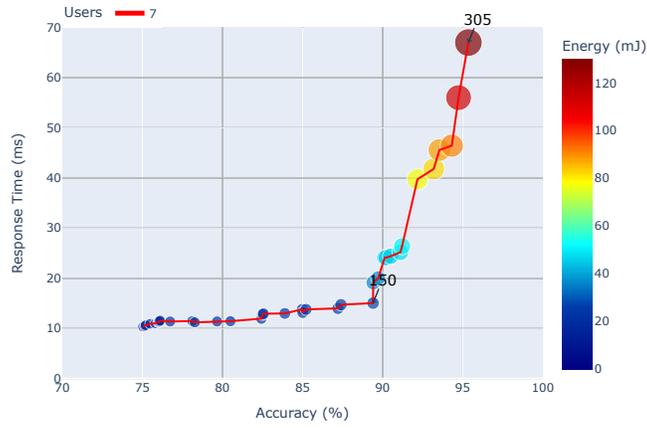

**Fig. 32.** Pareto-Front for User 7 with configurations with an accuracy greater than 75%. Some configurations highlighted (configuration ID is shown).

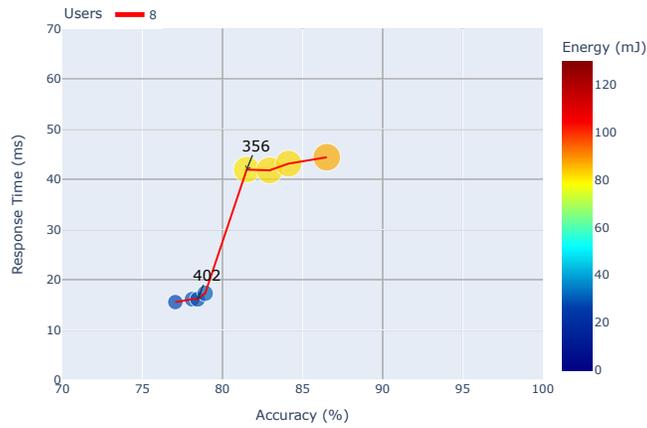

**Fig. 33.** Pareto-Front for User 8 with configurations with an accuracy greater than 75%. Some configurations highlighted (configuration ID is shown).

## B    Individual Users - Sampling Frequency

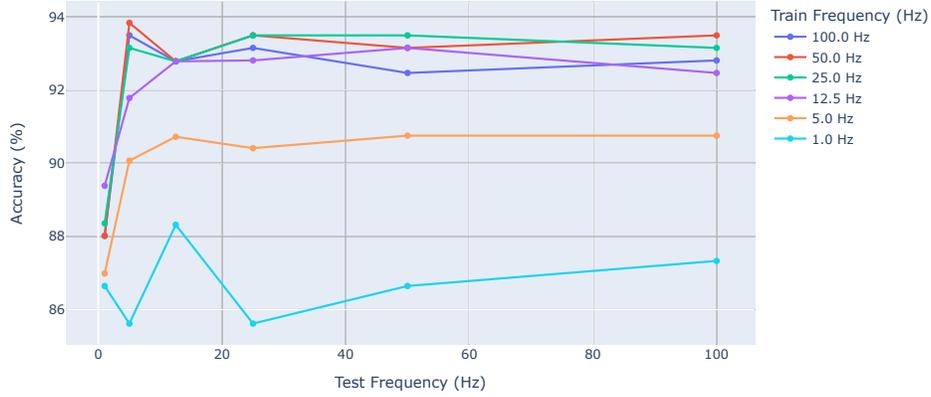

**Fig. 34.** Variation of the sampling frequency of the train and test sets for User 2, from 1 Hz to 100 Hz.

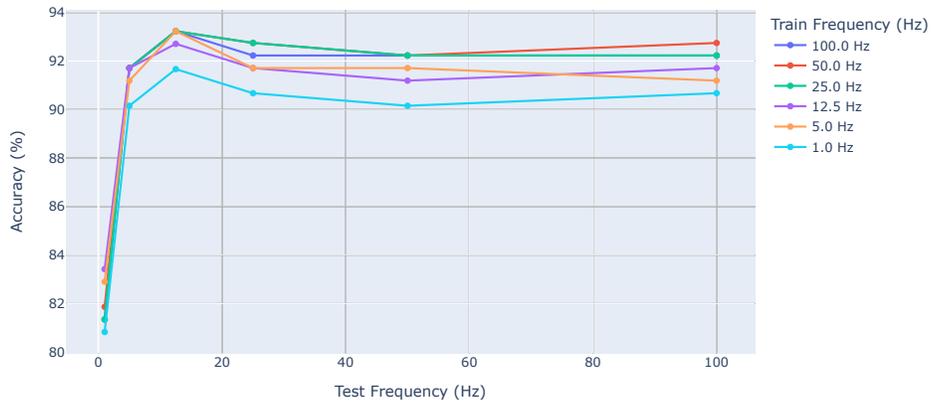

**Fig. 35.** Variation of the sampling frequency of the train and test sets for User 3, from 1 Hz to 100 Hz.

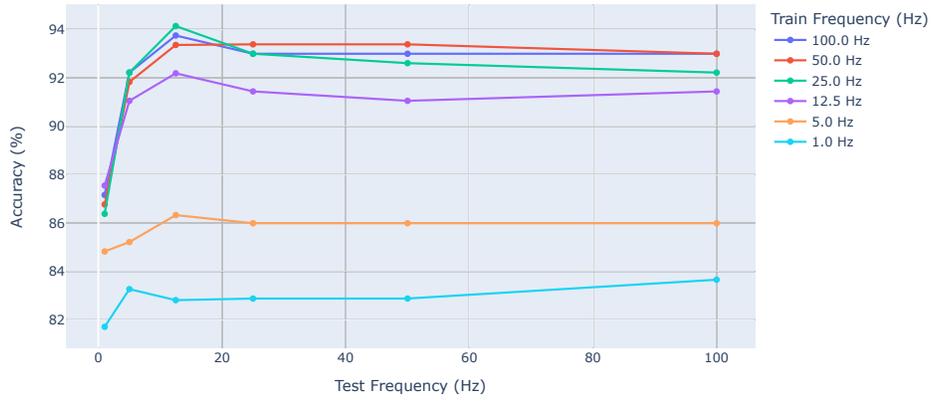

**Fig. 36.** Variation of the sampling frequency of the train and test sets for User 4, from 1 Hz to 100 Hz.

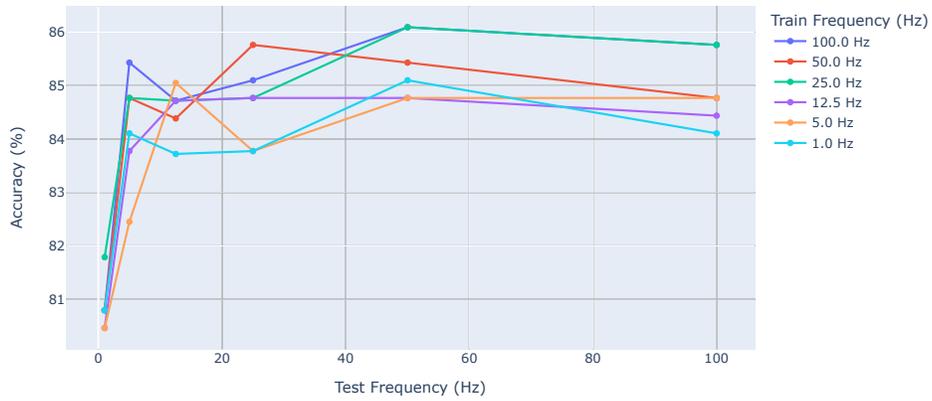

**Fig. 37.** Variation of the sampling frequency of the train and test sets for User 5, from 1 Hz to 100 Hz.

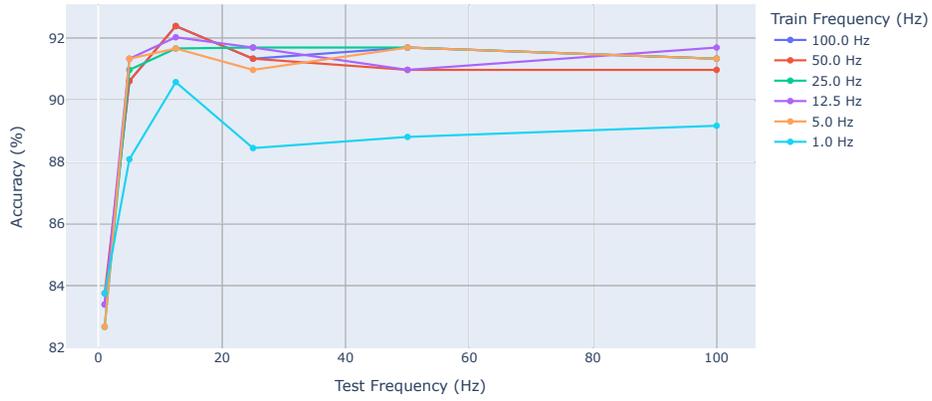

**Fig. 38.** Variation of the sampling frequency of the train and test sets for User 6, from 1 Hz to 100 Hz.

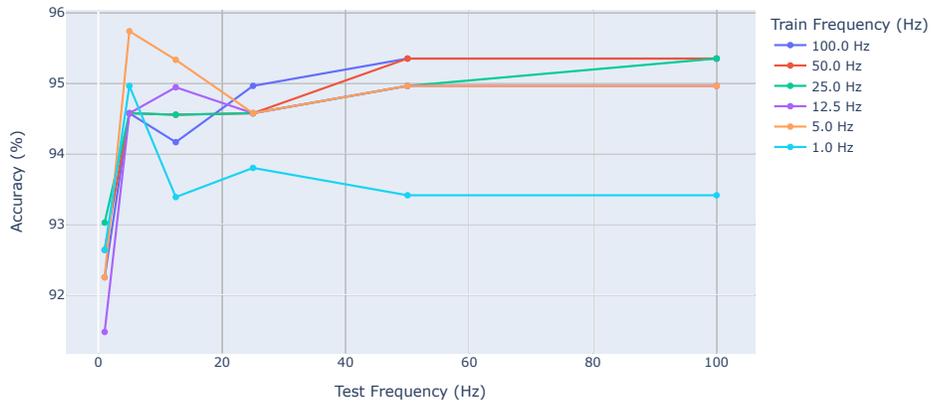

**Fig. 39.** Variation of the sampling frequency of the train and test sets for User 7, from 1 Hz to 100 Hz.

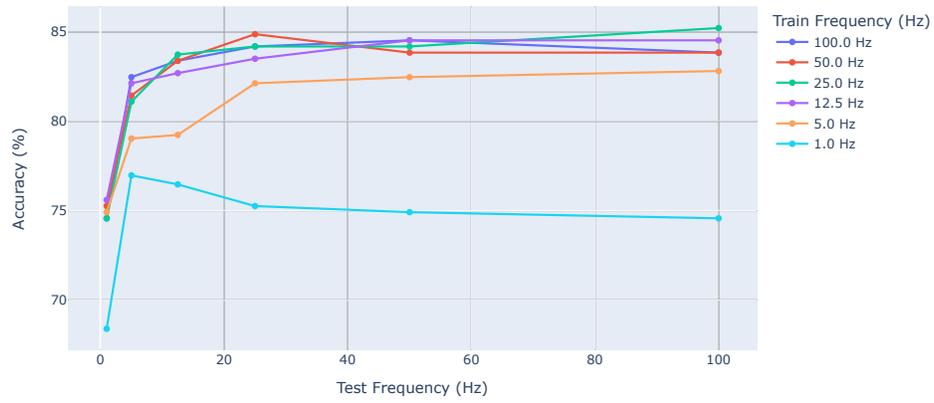

**Fig. 40.** Variation of the sampling frequency of the train and test sets for User 8, from 1 Hz to 100 Hz.



\end{document}